\documentclass[fleqn,usenatbib]{mnras}

\usepackage{newtxtext,newtxmath}

\usepackage[T1]{fontenc}
\usepackage{ae,aecompl}

\usepackage{graphicx}	
\usepackage{amsmath}	
\usepackage{amssymb}	

\usepackage{threeparttable}

\newcommand{\RM} [1]{\mathrm{#1}}

\newcommand{\EQ}[1]{equation~(\ref{eq:#1})}
\newcommand{\FIG}[1]{Fig.~\ref{fig:#1}}
\newcommand{\TAB}[1]{Table~\ref{tab:#1}}

\title[Non-detection of FRBs from six GRB remnants]{Non-detection of fast radio bursts from six gamma-ray burst remnants with possible magnetar engines}

\author[Yunpeng Men et al.]
{Yunpeng Men$^{1,2,3}$,
Kshitij Aggarwal$^{4,5}$,
Ye Li$^{1}$,
Divya Palaniswamy$^{3}$,\newauthor
Sarah Burke-Spolaor$^{4,5}$\thanks{E-mail: sarah.spolaor@mail.wvu.edu},
K. J. Lee$^{1}$\thanks{E-mail: kjlee@pku.edu.cn},
Rui Luo$^{1,2,8}$,
Paul Demorest$^{6}$,\newauthor
Shriharsh Tendulkar$^{7}$,
Devansh Agarwal$^{4,5}$,
Olivia Young$^{4,5}$,
Bing Zhang$^{3}$\thanks{E-mail: zhang@physics.unlv.edu}
\\
$^{1}$Kavli Institute for Astronomy and Astrophysics, Peking University, 
Beijing 100871, China\\
$^{2}$Department of Astronomy, School of Physics, Peking University, Beijing 100871, China\\
$^{3}$Department of Physics and Astronomy, University of Nevada Las Vegas, NV 89154, USA\\
$^{4}$Department of Physics and Astronomy, West Virginia University, Morgantown, WV 26505, USA\\
$^{5}$Center for Gravitational Waves and Cosmology, West Virginia University, Chestnut Ridge Research Building, Morgantown, WV, USA\\
$^{6}$National Radio Astronomy Observatory, Socorro, NM 87801, USA\\
$^{7}$Department of Physics and McGill Space Institute, McGill University, 3600 University Street, Montreal, QC H3A 2T8, Canada\\
$^{8}$CSIRO Astronomy and Space Science, Australia Telescope National Facility, Box 76, Epping, NSW 1710, Australia
}

\date{Accepted XXX. Received YYY; in original form ZZZ}

\pubyear{2019}

\begin{document}
\label{firstpage}
\pagerange{\pageref{firstpage}--\pageref{lastpage}}
\maketitle

\begin{abstract}
The analogy of the host galaxy of the repeating fast radio burst (FRB) source FRB 121102 and those of long gamma-ray bursts (GRBs) and super-luminous supernovae (SLSNe) has led to the suggestion that young magnetars born in GRBs and SLSNe could be the central engine of repeating FRBs. We test such a hypothesis by performing dedicated observations of the remnants of six GRBs with evidence of having a magnetar central engine using the Arecibo telescope and the Robert C. Byrd Green Bank Telescope (GBT). A total of $\sim 20$ hrs of observations of these sources did not detect any FRB from these remnants. Under the assumptions that all these GRBs left behind a long-lived magnetar and that the bursting rate of FRB\,121102 is typical for a magnetar FRB engine, we estimate a non-detection probability of $8.9\times10^{-6}$. Even though these non-detections cannot exclude the young magnetar model of FRBs, we place constraints on the burst rate and luminosity function of FRBs from these GRB targets.
\end{abstract}




\section{Introduction}
Fast radio bursts are bright millisecond radio pulses with dispersion measure (DM) in excess of the Galactic contribution along the line of sight \citep{Lorimer2007,Petroff2016}. At least two FRB sources (FRB\,121102 and FRB\,180814) are found to repeat \citep{Spitler2014apj, Scholz2016apj, CHIME2019ar}, suggesting a non-catastrophic progenitor system at least for some FRBs. FRB 121102 was localized in a host galaxy with a redshift $z = 0.19273(8)$ \citep{Tendulkar2017apj}, confirming the cosmological origin of FRBs.
Whether or not all FRBs are repeating sources is subject to debate \citep{Palaniswamy2018,Caleb2019}. In the literature, the FRB progenitor models can be grouped into two categories: the non-catastrophic models (related to repeating FRBs) such as giant magnetar flares \citep[e.g.][]{Kulkarni2014apj, Katz2016apj}, giant pulses from young pulsars or magnetars \citep[e.g.][]{Popov2010ar, Connor2016mn, Cordes2016mn, Murase2016, Kashiyama2017apj, Metzger2017apj, Margalit2018}, and interacting models \citep[e.g.][]{Zhang2017} among others; and the catastrophic models (related to non-repeating FRBs) such as 
collapse of supra-massive neutron stars \citep{Falcke2014aa, Zhang2014apj} and mergers of compact stars \citep[e.g.][]{Totani2013pa, Wang2016apj, Zhang2016apj}. See \cite{platts18} for a complete list of FRB progenitor models.

Gamma-ray bursts (GRBs) are most luminous explosions in the universe, signaling core collapse of massive stars or mergers of compact objects \citep{meszaros06,zhang18}. They are much rarer than FRBs. However, the following theoretical arguments have been made to suggest that a small fraction of FRBs could be associated with GRBs for different reasons: 1. A good fraction of both long and short GRBs have an X-ray plateau followed by a rapid decay, which are best interpreted as a supramassive neutron star collapsing to a BH at $\sim (10^2-10^4)$ s after the GRB triggers \citep{Troja2007,Rowlinson2010,Rowlinson2013,Lü2014,Lv2015}. If FRBs are produced from the so-called ``blitzar'' scenario \citep{Falcke2014aa}, an FRB would be produced with such a delay time after these GRBs \citep{Zhang2014apj}; 2. If a small fraction of non-repeating FRBs are related to mergers of binary neutron stars \citep{Totani2013pa}, it is possible that an FRB may be associated with a short GRB \citep[e.g.][]{Wang2016apj,Zhang2016apj};  3. If FRBs are produced when the magnetosphere of a neutron star is re-configured by an external astrophysical stream, a GRB could be the source of the astrophysical stream to trigger FRBs \citep{Zhang2017}; 4. Within the young magnetar model for repeating FRBs \citep[e.g.][]{Metzger2017apj,Margalit2018}, a GRB could be the progenitor of the young magnetar that later produce FRBs. Such a connection is promising in view of the similarity of the host galaxy of FRB\,121102 (a dwarf star forming galaxy at redshift $z = 0.19273(8)$, \citealt{Chatterjee2017nat,Marcote2017apj,Tendulkar2017apj}), to those of long GRBs and superluminous supernovae (SLSNe) \citep{Tendulkar2017apj,Metzger2017apj,Nicholl2017apj}. Moreover, \citet{Eftekhari2019apj} showed tentative evidence for a radio source coincident with the SLSNe PTF10hgi, similar to that of FRB\,121102 \citep{Chatterjee2017nat,Marcote2017apj,Tendulkar2017apj}, which could be produced by a magnetar central engine. Within such a picture, a GRB should proceed the repeating FRBs by months to decades. 

Searching for GRB-FRB associations have been carried out in the past to test the first three hypothesis so far with null results  \citep{Bannister2012,Palaniswamy2014,DeLaunay2016,Cunningham2019}. More dedicated searches for GRBs around the time of FRBs, and for FRBs following GRBs (especially short GRBs) within $<100$ s of the burst triggers are encouraged in order to rule out these possibilities. 

In this work, we test the fourth hypothesis of GRB/FRB associations, which is related to repeating FRBs. So far, only two repeating FRBs have been reported in the literature, FRB\,121102 \citep{Spitler2016nat, Scholz2016apj} and FRB\,180814 \citep{CHIME2019ar}. Searching for GRBs from archival Fermi data have been carried out for these two sources with null results \citep{Yamasaki2016,ZhangZhang17,Xi17,Yang19,Guidorzi19}. We adopt an opposite approach. We first identify several historical GRBs likely having a magnetar central engine based on  their lightcurves. We then perform dedicated searches for FRBs from these GRB remnants using the Arecibo telescope and the Robert C. Byrd Green Bank Telescope (GBT). The total search time is about 20\,hrs. No FRBs have been detected. We present the details of our observations and data analysis results are presented in Section 2 and 3, respectively. The implications of our results on FRB prgenitor systems are presented in Section 4, and the conclusions are summarized in Section 5.

\begin{table*}
    \centering
    \caption{The parameters of the observed target GRBs and observations.}
    \begin{threeparttable}
    \begin{tabular}{lllllllll}
    \hline
        GRB name$^\ast$ & Redshift & RA & Dec & $\RM{DM_{IGM}}$ & $\RM{DM_{MW}}^{\dagger}$ & Obs. telescope & Obs. time & Comments \\
        (yymmdd) &  & (h\,:\,m\,:\,s) & ($^{\circ}$\,:\,$\arcmin$\,:\,$\arcsec$) & ($\RM{cm^{-3}\,pc}$) & ($\RM{cm^{-3}\,pc}$) &  & (minutes) &  \\
    \hline
        030329 & 0.168 & 10:44:50.00 & +21:31:17.8 & 147 & 17 & Arecibo & 340.7 & LGRB+SN2003dh\\
        130603B & 0.3564 & 11:28:48.16 & +17:04:18.0 & 311 & 29 & Arecibo & 448.8 & short GRB\\
        111225A & 0.297 & 00:52:37.21 & +51:34:19.5 & 259.875 & 118.09 & GBT & 76.5 & LGRB\\
        051109B & 0.08 & 23:01:50.30 & +38:40:46.7 & 70.0 & 71.17 & GBT & 131.3 & LGRB\\
        111005A & 0.013 & 14:53:07.74 & -19:44:08.9 & 11.375 & 51.12 & GBT & 82.5 & LGRB\\
        980425 & 0.0085 & 13:25:41.93 & -26:46:55.7 & 7.43 & 53.59 & GBT & 70.6 & LGRB+SN1998bw\\
    \hline
    \end{tabular}
    $^\ast$ GRB targets with data damaged are not listed.\\
    $^\dagger$ $\RM{DM_{MW}}$ is Galactic contribution to the overall DM, which is estimated by integrating the NE2001 model \citep{Cordes2002} to the edge of the Galaxy.
    \end{threeparttable}
    \label{tab:GRBs}
\end{table*}

\section{Observations}

The observations were performed with the Arecibo and GBT. Each telescope was pointed at different GRB remnants multiple times. The parameters of target GRBs are listed in \TAB{GRBs}. These target GRBs were selected based on three criteria: (1) They are nearby GRBs with z < 0.4; (2) They fall into the declination ranges of the telescopes; (3) Four of them show an X-ray lightcurve plateau signature consistent with the magnetar central engine criteria as defined in \citet{Lü2014}. Two famous pre-Swift nearby long GRBs, GRB 980425 \citep{Galama1998} and GRB 030329 \citep{Hjorth2013}, are also selected. These two GRBs have Type Ic SN associations and have been suggested to be powered by magnetars \citep{Mazzali2014}. Among the six sources, one source, GRB 130603B, is a short GRB, whose X-ray lightcurve and the kilonova signature can be self-consistently interpreted as being powered by a magnetar central engine \citep{Fan2013apjl}. The newly localized FRB 180924 \citep{Bannister2019} has an environment consistent with that of a short GRB, justifying search for FRBs from short GRBs like GRB 130603B. The other five GRBs are all long GRBs. The ages of these GRBs range from 20 years (GRB 980425) to 5 years (GRB 111225A), which fall into the suggested magnetar age range that is favorable for FRB production and detection \citep{Metzger2017apj}.

\subsection{Arecibo}
Four GRB sources were observed for 25 one-hour durations at Arecibo, but unfortunately, most of the data were damaged in the Hurricane Maria and only about ten hours of data from two GRB targets (long GRB\,030329 and short GRB\,130603B)  are available, as shown in \TAB{GRBs}.

The observations were conducted with the single-pixel L-wide receiver. The central frequency of the observations is 1440 MHz and the bandwidth is 580 MHz. The system temperature $T_\RM{sys}$ is $30$ K and the gain $G$ is 10.5 $\RM{K\,Jy^{-1}}$. The minimum detectable flux is 
\begin{equation}
	S_\mathrm{min}=\beta \frac{\gamma\,T_\mathrm{sys}}{G\,\sqrt{\mathrm{BW}\,\tau\,N_\mathrm{p}}}\,,
	\label{eq:sensitivity}
\end{equation}
where $\gamma$ is the threshold of signal-to-noise ratio (S/N), $\beta\simeq 1$ is the digitisation factor, $\mathrm{BW}$ is the bandwidth, $N_\mathrm{p}$ is the number of polarisations, and $\tau$ is the pulse width. By choosing 3 ms as the reference width of a possible FRB, the flux thresholds for S/N>7 is 10.7 mJy. The data were recorded with the PUPPI backend in incoherent search mode with a 800 MHz total bandwidth, 2048 frequency channels, 256 $\mu s$ time resolution, and the summed polarization.

\subsection{Green Bank Telescope}

Four long GRB sources (GRBs 980425, 052209B, 111005, and 111225) were observed at GBT, as listed in \TAB{GRBs}. The observational times of the target GRBs are listed in \TAB{GRBs}.

The observations were performed with the 820 MHz and 2 GHz receivers. The system temperature is $T_\RM{sys}$ of 25 K and the gain is 2 $\RM{K\,Jy^{-1}}$, which yields the minimum detectable flux of 38.3 mJy according to \EQ{sensitivity}. We recorded the observation data with the GUPPI backend in the search mode, in which nominal 800 MHz bands are sampled with 256 $\mu$s time resolution and 2048 spectral channels. The data are polarization summed.

\section{Data Reduction and Results}

The observation data are processed with the pipeline 'Burst Emission Automatic Roger' (BEAR) which is used to search for dispersed burst signals \citep{W1}. The raw data are first normalized with zero mean and unit variance in each frequency channel. We perform the radio frequency interference (RFI) mitigation using the \emph{zero-DM matched filter}, which estimates the waveform of the zero DM signal and subtracts only the corresponding contribution from each frequency channel \citep{W1}. The data are then dedispersed with trial DMs ranging from 0 to 1000 $\RM{cm^{-3}\,pc}$ with a DM step 0.1 $\RM{cm^{-3}\,pc}$. The maximum DM searched is roughly twice as high as the estimated IGM and Milky Way contributions to DM listed in \TAB{GRBs}. The DM step gives a smearing compared to the time resolution. We then search for burst signals in the dedispersed time series using the \emph{matched filter} that convolves the dedispersed time series with a series of boxcar matched filters with a geometric series of width covering 0.256-20 ms. The burst candidates with S/N$>7$ are saved, while the duplicated candidates are removed using the candidate clustering algorithm (for a general discussion, see \citet{W1}).

Before reducing the data from the GRB remnants, we test the search pipeline with the test observation data of PSR\,B1133+16 at Arecibo. All the single pulses of PSR\,B1133+16 can be detected with BEAR. After the burst signal search in the data, about 23000 and 900 candidates were reported by BEAR in the Arecibo and GBT data, respectively. We scan all the candidates with eyes and they are all recognized as RFI signals. As a result, we have no positive detection of celestial bursts in the data.

The data were also processed using Graphics Processing Unit (GPU) accelerated transient detection pipeline {\sc heimdall}\footnote{https://sourceforge.net/projects/heimdall-astro/} \citep{barsdell12}. The parameters used in the search were: S/N $>$ 6, 20 $\leq$ DM $\leq$ 10,000~pc~cm$^{−3}$ and width $\leq$32~ms. This resulted in 27,779 candidates which were then classified using model \texttt{a} of the deep learning based classifier {\sc fetch}\footnote{https://github.com/devanshkv/fetch} \citep{Agarwal19} to identify FRBs from RFI. It identified 305 candidates as potential FRBs, which were then visually inspected and found to be RFI.

\begin{figure}
    \centering
    \includegraphics[width=\columnwidth]{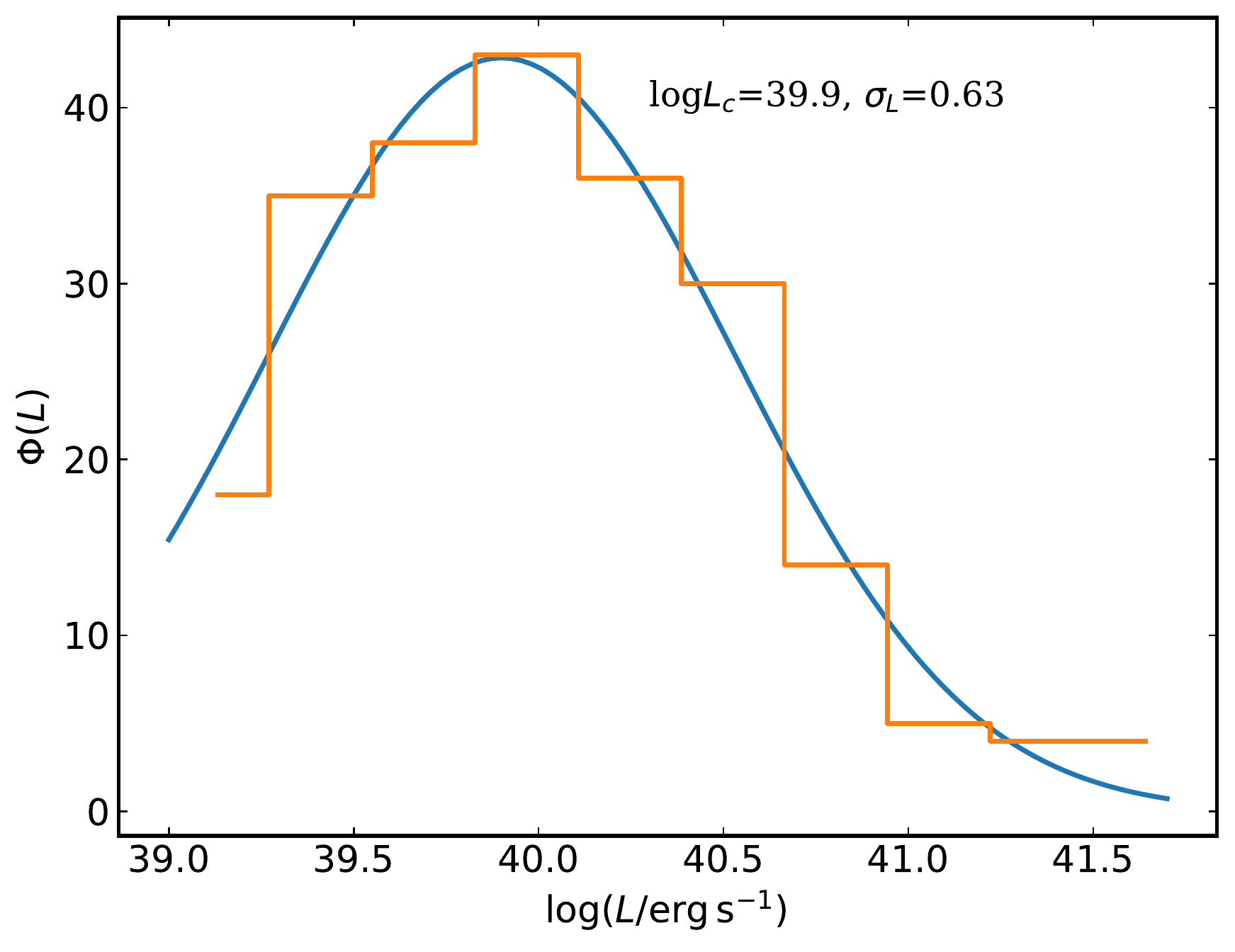}
    \caption{The luminosity distribution of the bursts of FRB\,121102 reported currently. The luminosity are derived using \EQ{luminosity} with $z=0.19273$. We choose a characteristic bandwidth of 200 MHz \citep{Gourdji2019apj}. The best fitting using \EQ{luminosity_function} gives $\mathrm{log}L_c=39.9$ and $\sigma_L=0.63$}
    \label{fig:luminosity}
\end{figure}

\section{Testing the young magnetar model for FRBs}

\subsection{Non-detection probability}
\label{sec:non-detection}

The non-detection of FRBs from these GRB remnants can be used to test the young magnetar -- FRB association hypothesis. 

We can estimate the non-detection probability from our observations based on three assumptions: 1) FRB\,121102 burst rate is typical for repeating FRB sources that host a young magnetar; 2) The distribution of the number of bursts $n$ in a given duration $T$ is poissonian, i.e.
\begin{equation}
    P(n| T, \eta)=\frac{(\eta\,T)^n\,e^{\eta\,T}}{\Gamma(n+1)}\,, 
\end{equation}
where $\eta$ is the intrinsic burst rate; 3) the luminosity function of the bursts $\Phi(L)$ for an FRB source is Gaussian logarithm, i.e.
\begin{equation}
    \Phi(L|L_c, \sigma_L)=\frac{1}{\sqrt{2\pi}\sigma_L\,L}e^{-\frac{1}{2}\left[\RM{log}\,\left(\frac{L}{L_c}\right)/\sigma_L\right]^2}\,,
    \label{eq:luminosity_function}
\end{equation}
where $L_c$ is the peak luminosity and $\sigma_L$ is the standard deviation of $\RM{log}\,L$. This last assumption is based on the luminosity distribution statistics of 227 bursts detected from FRB\,121102 \citep{Spitler2014apj, Spitler2016nat, Scholz2016apj, Scholz2017apj, Law2017apj, Hardy2017mn, Spitler2018apj, Michilli2018nat, MAGIC2018mn, Zhang2018apj, Hessels2019apj, Gourdji2019apj}, as shown in \FIG{luminosity}.

With the above assumptions, the non-detection probability of FRBs from these targets can be written as
\begin{equation}
    P_{\RM{non}}(\eta, L_{\rm c}, \sigma_L) = 1 - P(n>0|T,\eta) \int_{L_\RM{min}}^{\infty} \Phi(L|L_c,\sigma_L) d L\,,
    \label{eq:non-detection probability}
\end{equation}
with
\begin{equation}
    L_\RM{min} = 4\pi\,r_\RM{L}^2(z)\,\RM{BW} S_\RM{min}\,.
    \label{eq:luminosity}
\end{equation}
The luminosity distance $r_\RM{L}(z)$ at the redshift $z$ is defined as
\begin{equation}
    r_{\rm L}(z)=\frac{c(1+z)}{H_{\rm 0}}\int _0^{z}\frac{1}{E(z)} dz\,,
\end{equation}
where the Hubble constant $H_0 = 67.8\ {\rm km\,s^{-1}\,Mpc^{-1}}$ 
\citep{Planck2018ar}, and $c$ is the light speed. $E(z)$ is the logarithmic 
time derivative of the cosmic scale factor, which is defined as
\begin{equation}
    E(z)=\sqrt{\Omega_{\rm m}\,(1+z)^3+\Omega_{\Lambda}}\,,
\end{equation}
where the total matter density $\Omega_{\rm m}=0.308$ and the dark energy density $\Omega_{\Lambda}=0.692$ have been adopted \citep{Planck2018ar}.

We conservatively determine the parameters of FRB\,121102 using the data in \citet{Li2019ar}, who summarized the burst data of FRB\,121102 and exhibited the statistical properties of the bursts. We adopt the observed burst rate $\eta=3\ \RM{hr^{-1}}$ as the intrinsic rate. The two luminosity function parameters $L_{\rm c}=10^{39.9}\,\RM{erg/s}$ and $\sigma_L=0.63$ are estimated through the flux density distribution with $z=0.19273$ for FRB\,121102, as shown in \FIG{luminosity}. Using \EQ{non-detection probability}, the non-detection probability can be determined for each GRB target with the observational time and redshift in \TAB{GRBs}. The total non-detection probability of all targets is the product of each one, which is $8.9\times10^{-6}$.

Such a low probability may be regarded as evidence disfavoring the possibility that GRB remnants with a young magnetar engine as the source of repeating FRBs. However, this possibility cannot be ruled out by the data. Possibilities to lower the non-detection probability include: 1) Not all these GRB remnants harbor a magnetar (i.e. the arguments in favor of a magnetar engine are for some reasons unjustified); 2) Maybe some of the young magnetars formed the remnants are supramassive and have collapsed to black holes when our observations started; 3) The supernova ejecta may not yet become transparent to free-free emission at the observing frequency (\citet{Margalit2018mn} showed that the timescales for free-free transparency of the ejecta can vary from decades to up to a 100 years); 4) The synchrotron self-absorption by a flare-powered nebula may dominate the radiative processes; 5) FRB\,121102 bursts are sporadic, and there are quiescent period \citep[e.g.][]{Price2018rn}; 6) FRB\,121102 source is abnormally active compared with most young magnetars, as have been also suggested in previous studies \citep{Palaniswamy2018,Caleb2019}. 

\subsection{Constraints on the intrinsic burst rate and luminosity function}
\label{sec:constraints}

In the following we assume that the non-detection is caused by the last two reasons discussed above and constrain the intrinsic FRB burst rate and luminosity function of young magnetars. By doing so, we have assumed that all these GRB remnants harbor young magnetars and they are still alive at the time of our observations. We have also made the assumptions 2) and 3) made in section \ref{sec:non-detection}.

\begin{figure*}
    \centering
    \includegraphics[width=\columnwidth]{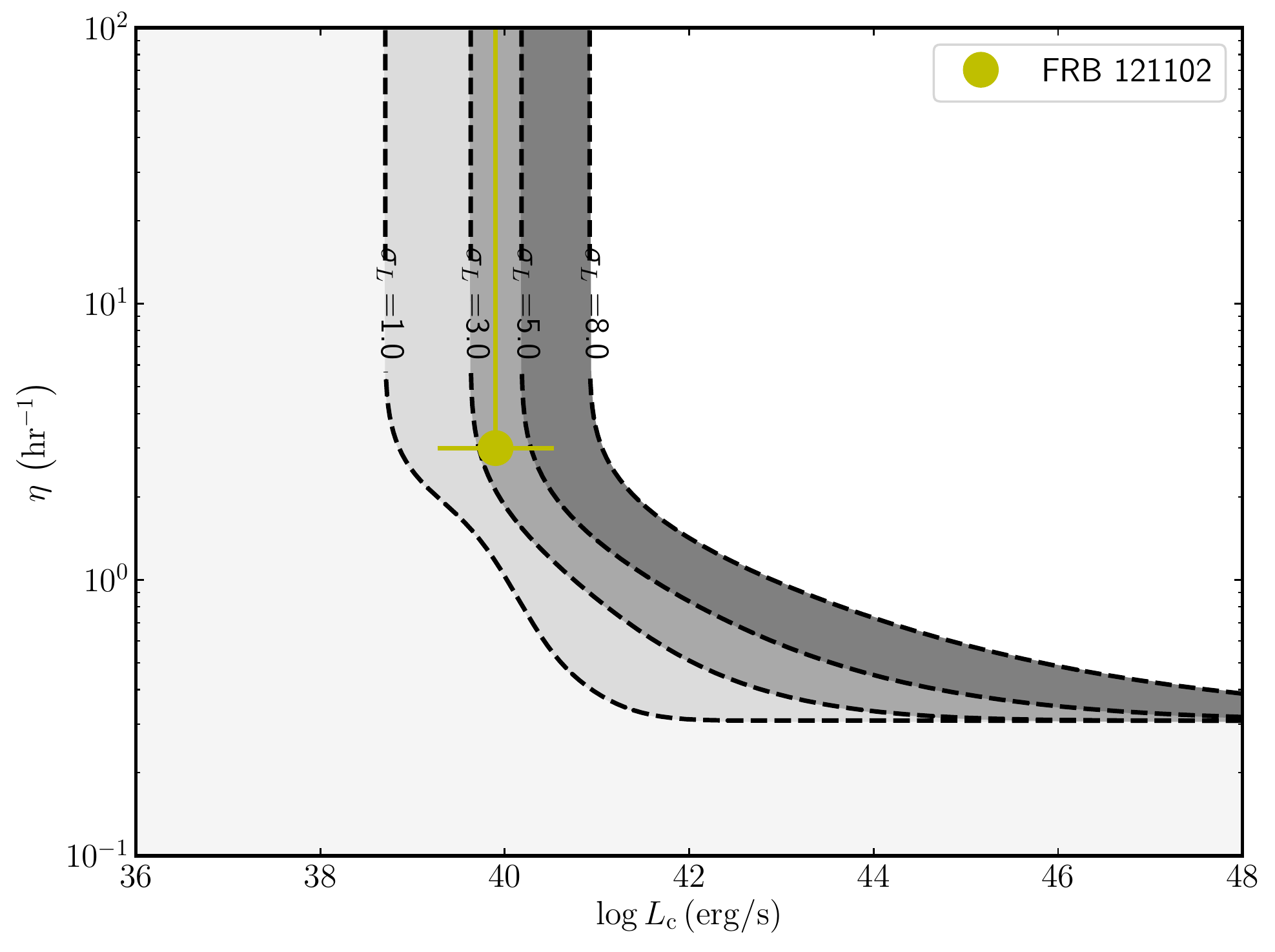}
    \includegraphics[width=\columnwidth]{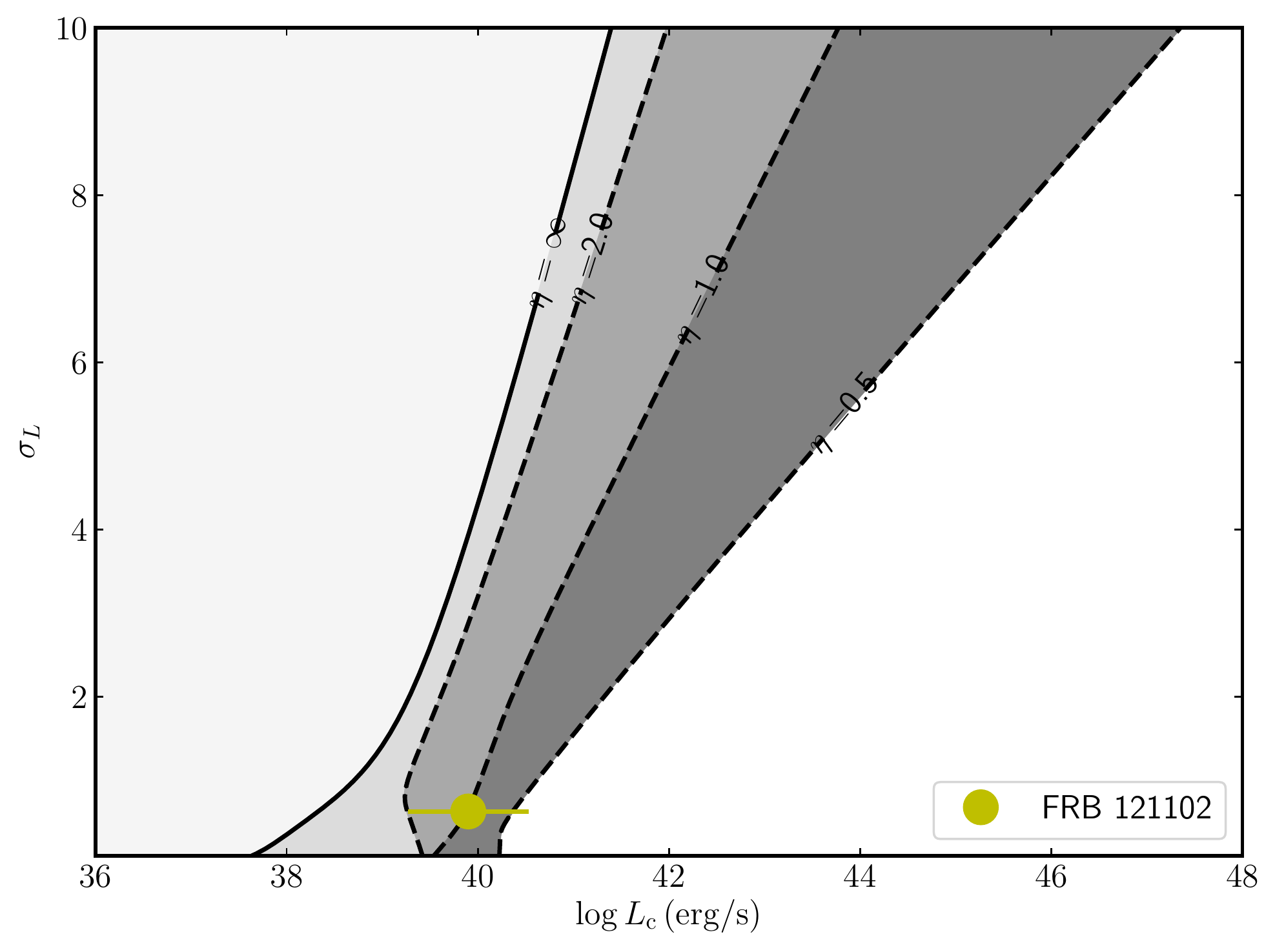}
    \caption{Constraints on the intrinsic burst rate $\eta$, the peak luminosity $L_{\rm c}$ and the standard deviation of luminosity function $\sigma_L$ based on the non-detection results with the total non-detection probability of 0.0027. The brighter region allows the existence of parameters. The FRB\,121102 with $\eta=3\,\RM{hr^{-1}}$, $L_{\rm c}=10^{39.9}\,\RM{erg/s}$ and $\sigma_L=0.63$ is plotted in the figures . Left panel: The constrain region of $\eta$ and $L_{\rm c}$ with $\sigma_L=1, 3, 5, 8$. Right panel: The constrain region of $\sigma_L$ and $L_{\rm c}$ with $\eta=0.5, 1, 2, \infty$.}
    \label{fig:constraints}
\end{figure*}

Using \EQ{non-detection probability}, the constraints on the intrinsic burst rate and luminosity function can be inferred by choosing the non-detection probability of 0.0027, which equals to 3$\sigma$ significance. Here we give the $\eta-L_{\rm c}$ curve with $\sigma_L=1, 3, 5, 8$ and $\sigma_L-L_{\rm c}$ curve with $\eta=0.5, 1, 2, \infty$. The results are shown in \FIG{constraints}, in which the allowed regions are to the left of the contours. We also place FRB\,121102 with $\eta=3\ \RM{hr^{-1}}$, $L_{\rm c}=10^{39.9}\ \RM{erg/s}$ and $\sigma_L=0.63$ in the plot. More extensive searches of FRBs from similar targets in the future can tighten these constraints further.

\section{Conclusions and discussion}

We have performed dedicated searches of FRBs from six nearby historical GRB sources (5 long and 1 short) which are consistent with having a millisecond magnetar central engine. The motivation is to test the hypothesis that these young magnetars born from GRB remnants can be the sources of repeating FRBs, as suggested by the similarity between the host galaxy of FRB\,121102 and those of long GRBs \citep{Tendulkar2017apj,Metzger2017apj,Nicholl2017apj}. Our 20 hr observations with Arecibo and GBT did not detect any FRB from these sources.

Under the assumptions that these sources indeed harbor magnetars similar to the putative magnetar that is powering FRB\,121102, that the burst rate is Poissonian, and that the luminosity function is Gaussian logarithmic, we conservatively estimated the non-detection probability of FRB\,121102-like bursts to be $8.9\times10^{-6}$. This result challenges the young magnetar scenario to power FRBs, even though the scenario cannot be ruled out. Assuming that the young magnetars still exist in these remnants and that they indeed produce FRBs without free-free absorption or synchrotron self-absorption, we place the constraints on the burst rate and luminosity function based on the non-detection observations.

Recent localizations of FRB\,180924 \citep{Bannister2019} and FRB\,190523 \citep{Ravi2019} suggest that the host galaxy of FRB\,121102 is not the norm of FRB hosts. A survey of possible host galaxies of nearby (excess DM < 100) FRBs also suggests that the FRB\,121102 host is abnormal \citep{Li2019}. All these suggest that LGRBs and SLSNe are likely not the dominant channel to produce FRB sources, which is consistent with the finding of our paper.

\section*{Acknowledgements}

YPM, KJL, and RL were supported by NSFC U15311243, National Basic Research Program of China, 973 Program, 2015CB857101 and XDB23010200. S. B.-S\ and K. A\ acknowledge support from NSF grant AAG-1714897. DA acknowledges support from the NSF awards OIA-1458952 and PHY-1430284. We are grateful to L.J. Zhao and D.Y. Chen for helpful discussion.




\bibliographystyle{mnras}
\bibliography{ms} 


\bsp	
\label{lastpage}
\end{document}